\documentclass[fleqn,usenatbib]{mnras}

\usepackage{newtxtext,newtxmath}
\usepackage{amsmath,marvosym}
\usepackage{multirow}
\usepackage{xcolor}
\usepackage{ulem}
\usepackage{booktabs}
\usepackage{multirow}
\usepackage{subcaption}
\usepackage[T1]{fontenc}
\usepackage[british]{babel}
\DeclareRobustCommand{\VAN}[3]{#2}
\let\VANthebibliography\thebibliography
\def\thebibliography{\DeclareRobustCommand{\VAN}[3]{##3}\VANthebibliography}

\usepackage{graphicx,epsfig}	
\usepackage{amsmath}	
\usepackage{tipa}

\title[]{Dust Colour, Phase Behaviour, and Monte Carlo Modelling of Interstellar Comet 3I/ATLAS from 4 au Pre‑ to 4 au Post‑Perihelion}

\author[F. Moreno et al.]{
Fernando Moreno,$^{1}$\thanks{E-mail: fernando@iaa.es}
Miquel Serra-Ricart,$^{2,3,4}$
Javier Licandro,$^{3,4}$
Pedro J. Guti\'errez,$^{1}$ 
\newauthor
Luisa M. Lara,$^{1}$
Irene Mariblanca-Escalona,$^{1}$
and Miguel R. Alarcon $^{2,3,4}$
\\
$^{1}$Instituto de Astrofísica de Andalucía, CSIC, Glorieta de la Astronomía s/n, 18008 Granada, Spain \\
$^{2}$Light Bridges, Observatorio Astronómico del Teide. Carretera del Observatorio del Teide, s/n, Güímar, Tenerife, Spain \\
$^{3}$Instituto de Astrof\'\i sica de Canarias, C/Vía Láctea s/n, 38205, La Laguna, Tenerife, Spain, and \\ 
$^{4}$Departamento de Astrofísica, Universidad de La Laguna, Avda. Astrofísico Francisco Sánchez, 38206 La Laguna, Tenerife, Spain \\
}

\date{Accepted XXX. Received YYY; in original form ZZZ}

\pubyear{2026}

\begin{document}
\label{firstpage}
\pagerange{\pageref{firstpage}--
\pageref{lastpage}}
\maketitle

\begin{abstract}

We report multi-band photometric imaging 
observations and Monte Carlo dust tail modelling of the interstellar comet 3I/ATLAS covering a wide range of heliocentric distances, from about 4~au pre-perihelion to 4~au post-perihelion. The extensive imaging data set allowed us to constrain the dust physical properties, ejection speeds, and production rates as a function of heliocentric distance. The post-perihelion observations, obtained at high cadence in multiple photometric bands (SDSS $g$, $r$, $i$, and luminance filters) and spanning phase angles between approximately $0.7^\circ$ and $30^\circ$, enabled us to determine the dust color and phase function. The resulting phase curve exhibits a prominent backscattering enhancement, distinct from those derived for Solar System comets, with an opposition surge of 0.1--0.4~mag, a width of 1--3$^\circ$, and a linear phase coefficient of 0.02--0.04~mag~deg$^{-1}$, consistent with independent pre-perihelion estimates. A possible interpretation of the imaging data, together with independent photometric measurements, indicates a dust size distribution characterized by a power-law index $\kappa \approx -3.5$, with minimum and maximum particle radii of $r_{\min} = 10~\mu\mathrm{m}$ and $r_{\max}$ in the interval 1--10~cm. 
The reported water production rate correlates well with the dust production rate post-perihelion, but fails to do so pre-perihelion, an effect possibly linked to the high CO$_2$/H$_2$O ratio measured before perihelion. The derived maximum dust-loss rate at perihelion is $(0.5$--$1.8)\times 10^{4}$~kg~s$^{-1}$.

\end{abstract}




\section{Introduction} 
\label{sec:Introduction}
Interstellar comet 3I/ATLAS, was discovered on 2025 July 1 
\citep{2025MPEC....N...12D} by the Asteroid Terrestrial- impact Last Alert System \citep[ATLAS,][]{2018PASP..130f4505T}, at a heliocentric distance of 4.51 au. Shortly after its discovery, \cite{2025ATel17264....1A}
reported cometary activity detected in deep $g$'-band images obtained on 2025 July 2. Precovery observations by TESS satellite show that the object was already active on 2025 May 7 
\citep{2025ApJ...991L...2F}, at heliocentric distance of 6.4 au, i.e.,  180 days before perihelion. 
This object represents a rare opportunity to investigate the physical properties of dust originating beyond the Solar System. As only the third confirmed interstellar object, its hyperbolic orbit and high inbound velocity firmly establish an extrasolar origin, providing a unique benchmark for comparing dust production and processing in exoplanetary environments with those observed in Solar System comets. Recent analyses of its discovery and orbital parameters confirm an eccentricity of $e\approx 6.1$ and a velocity at infinity of $v_{\infty}\approx$ 58 km s$^{-1}$  \citep{2025ApJ...989L..36S}, values consistent with an origin in the interstellar medium rather than the Oort Cloud. In fact, the exceptionally high D/H ratio reported for this comet---more than 30 times higher than that of Solar System comets---suggests that it formed in an environment significantly colder than the solar nebula at the time of the Sun's birth \citep{2026arXiv260307026S}. In addition to the high D/H, \cite{2026arXiv260306911C} 
 also reported high isotopic abundances of $^{12}$C/$^{13}$C relative to the values found in the Solar System, pointing to a cold environment as well. Notwithstanding this, the abundances of other molecules—such as CO, H$_2$CO, CH$_3$OH, and CH$_3$CN—relative to HCN do not appear anomalous, instead falling within the upper end of the range measured in Solar System comets \citep{2026arXiv260323240B}.

Early imaging and spectroscopic observations 
\citep[e.g.][among others]{2025ATel17264....1A,delaFuenteMarcos2025Assessing, 2025ApJ...990L...2J,2025A&A...700L..10A}
reveal that 3I/ATLAS exhibits a well‑developed coma with prominent dust structures, including extended filaments detected by spacecraft imaging campaigns. Data obtained by ESA’s JUICE mission show dust-rich outflows and complex morphological features in the inner coma, suggesting active dust release and possible fragmentation processes as the comet approached perihelion in late 2025 October \citep{Tubiana2026}. These observations highlight the importance of dust as a tracer of the comet’s thermal history and internal structure.

The dust environment of 3I/ATLAS is of particular interest because interstellar comets may preserve primordial solids formed in protoplanetary disks around other stars. Their dust composition, grain-size distribution, and ejection behaviour can therefore provide constraints on the diversity of solid material produced in extrasolar systems. The unusual chemical richness reported in early studies of 3I/ATLAS 
\citep{2025ApJ...991L..43C,2025MNRAS.542L.139B,
2025ApJ...989L..36S,
2025ApJ...990L...2J,2025ApJ...991L...2F}, combined with its active dust production, suggests that its nucleus may contain volatile-rich grains and refractory material shaped by physical conditions distinct from those of the early Solar System. Understanding the dust properties of 3I/ATLAS is thus essential for interpreting its activity and for placing it in the broader context of interstellar small bodies.

In this work, we present a detailed analysis of the dust environment of comet 3I/ATLAS, using ground-based imagery and light curves encompassing  heliocentric distances  from about 4 au pre-perihelion to 4 au post-
perihelion, with modelling of the dust properties (speed, size distribution, and production rates). Our aim is to characterize the physical properties of the dust population and to assess the implications for the formation and evolution of solids in extrasolar planetary systems, providing clues about the governing physical conditions. The European Space Agency's Comet Interceptor mission \citep{2024SSRv..220....9J} is primarily focused towards the first-ever in-situ characterization of a "dynamically new" comet or an interstellar object 
to characterize its surface properties, shape, structure, and the composition of its gas coma. Thus, studies of interstellar comets are of the utmost interest in preparation for that mission. We are particularly interested in dust properties, and to that end, the dust tail images are interpreted using a Monte Carlo dust tail code, called \texttt{COMTAILS}, which has recently been made publicly available to the scientific community \citep{2025A&A...695A.263M}.

\begin{figure*}
\includegraphics[angle=-90,trim=1cm 1cm 0cm 1cm,clip,width=19cm]{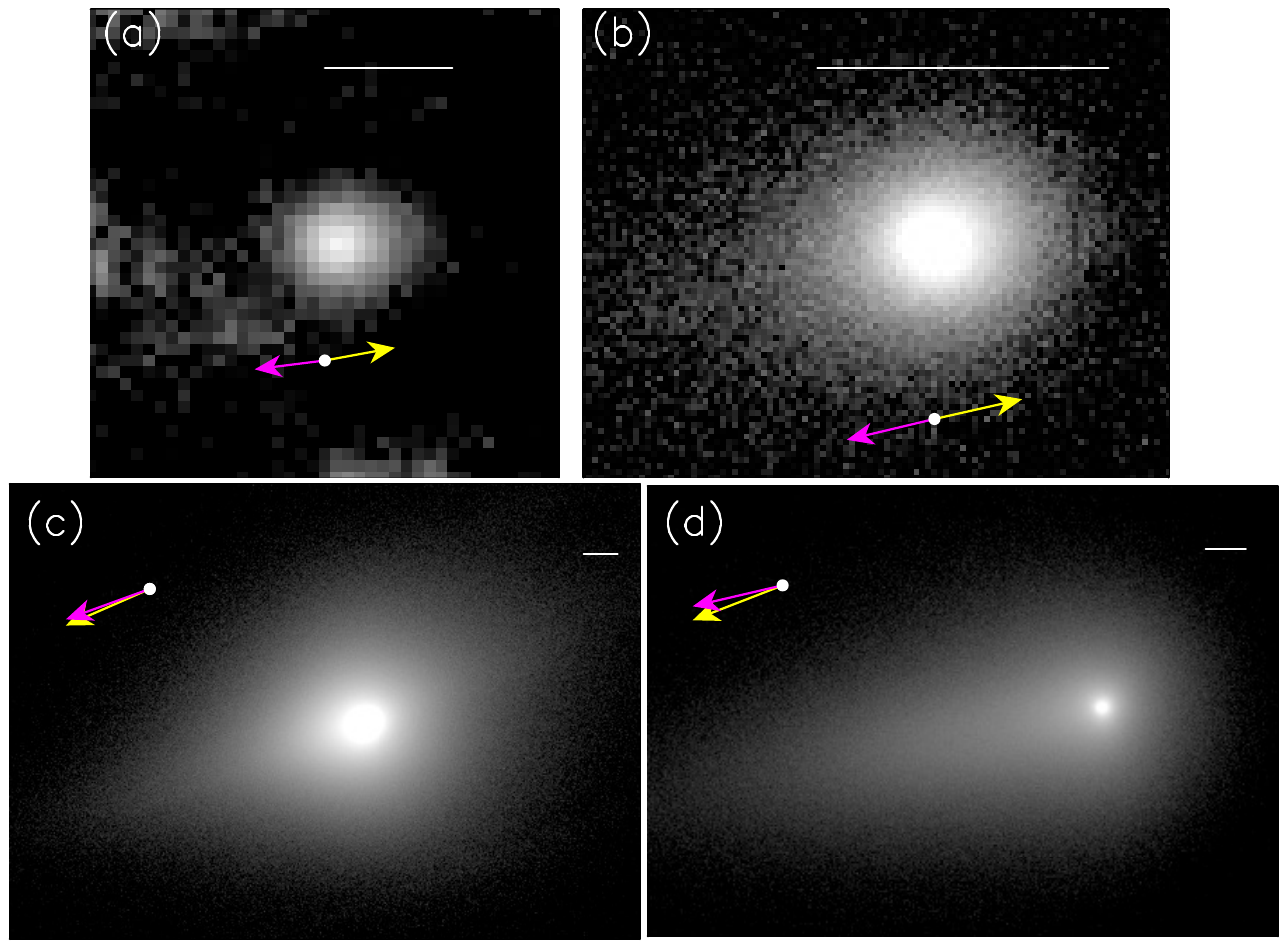}   
\caption{Representative images acquired on various ground-based observatories on different dates: (a) on 2025/07/20, using CAFOS at 2.2m telescope of the Centro Astron\'omico Hispano-Alem\'an, (b) on 2025/08/28, using ALFOSC at the 2.5m Nordic Optical Telescope, (c) on 2025/12/02 using the TST telescope at Teide Observatory in Tenerife, and (d) on 2026/01/11, also using the TST telescope. The bar near the top right on the images represents 20,000 km projected on the sky at the comet distance. The yellow arrows point to the Sun, and the magenta arrows point to the negative of the comet heliocentric velocity vector. All images are oriented North up, East to the left.}
\label{fig:example_figure}
\end{figure*}

\section{Observations}  \label{sec:Observations}

Pre-perihelion images were acquired with the 
Calar Alto Faint Object Spectrograph (CAFOS)  \citep{Meisenheimer1998} mounted at the Cassegrain focus of the 2.2m telescope at Calar Alto Observatory in southern Spain\footnote{\url{https://www.caha.es}}, and with the Alhambra Faint Object Spectrograph and Camera (ALFOSC) at the Nordic Optical Telescope (NOT)  \citep{2010ASSP...14..211D} at the Roque de los Muchachos Observatory at La Palma, Spain  \footnote {\url{https://www.iac.es/en/observatorios-de-canarias/roque-de-los-muchachos-observatory}}. The log of the pre-perihelion observations is given in Table \ref{tab:logobs_pre}. The images were acquired using Cousins R filters with 50 to 60 seconds exposure times. Reduction was performed using standard techniques (bias and flat-field correction) and calibrated using field stars from the Pan-STARRS1 (PS1) catalogue \citep{2012ApJ...750...99T}.  A median stack from the available images was produced each night. Image scales were 0.530 and 0.214 \arcsec~px$^{-1}$ at CAFOS and ALFOSC, respectively.

The post-perihelion images were acquired with
commissioning time of the Transient Survey Telescope
(TST\footnote{\url{https://tst.iac.es}}, MPC code Y64) at the Teide Observatory in Tenerife, Canary Islands, Spain. This telescope is a new 1‑m, wide field-of-view, fully robotic facility designed specifically for time‑domain surveys and rapid‑response observations.

The prime focus camera FERVOR-L (\textit{Fast Embedded-sCMOS Robotic Visible Observatory for Rapid transients}), based on sCMOS sensor (Sony IMX411, \citealt{Alarcon2023}) was used.
FERVOR-L has a pixel scale of 0.60 arcsec~px$^{-1}$ and an impressive field-of-view (FoV) $2.4\times1.8$ deg$^2$ ($14304\times10748$ pixels). 

All raw frames were corrected for bias and twilight flat-field variations using standard procedures. A WCS (world coordinate system) solution was obtained using Astrometry.
net \citep{2010AJ....139.1782L} against Gaia DR2 \citep{GaiaDR2} and all comet
images have been cropped around the comet’s optocenter (final FoV of 10 arcmin).

  The images were obtained through the SDSS $g$, $r$, and $i$ filters, as well as a luminance ($Lum$) filter, which has an approximately constant spectral response from 400 to 720\,nm and blocks radiation outside that interval. Exposure times were set to 40\,s, except for the $i$-band images, which were exposed for  45\,s. Absolute calibration was performed using stars from the PS1 survey catalogue
  \citep{2012ApJ...750...99T}. The luminance filter images were calibrated assuming that the magnitudes can be computed as $L' = 0.5159\,g' + 0.4841\,r'$
where $g'$ and $r'$ are the corresponding PS1 catalogue magnitudes. This expression is valid for solar-type stars and provides a reliable photometric reference for the $Lum$ band.
  

 The object was observed continuously during 2025 December and 2026 January and February, covering the outbound arc from 1.8 to 4.3 au. Table \ref{tab:logobs_post} gives a summary of the observing log for those images for $g$,$r$, and $Lum$ filters. The $i$ filter was used during the observing dates in Table  \ref{tab:logobs_post} in 2026 January and February only. The total number of images analyzed were 2150 in the $g$ and in the $r$ filter, 675 in the $i$ filter, and 7930 in the $Lum$ filter.  

To retrieve the dust parameters, we used our Monte Carlo dust-tail code. As input, we considered all pre-perihelion images together with a selected subset of post-perihelion images obtained on the dates listed in Table~\ref{tab:sample_dates_post}. Figure~\ref{fig:example_figure} shows representative pre- and post-perihelion images, illustrating the diversity of observed morphologies. At large inbound heliocentric distances, the comet exhibits a compact coma with a short sunward-pointing tail (panel (a) in Figure~\ref{fig:example_figure}).  As the comet approached perihelion, its morphology evolved into an increasingly typical appearance, with the dust tail progressively aligning in the anti‑sunward direction (panel (b) in Figure \ref{fig:example_figure}). Once the comet became unobservable from the ground and had passed perihelion, it developed an increasingly prominent anti‑tail (panels (c) and (d) in Figure \ref{fig:example_figure}).
  
In addition to the images, for modelling purposes, we used the $R$-band magnitude data obtained by the amateur association \texttt{Cometas\_Obs}
\footnote{\url{http://www.astrosurf.com/cometas-obs/}}, which provide good coverage during both the inbound and outbound branches, as well as the precovery TESS magnitude data provided by \cite{}.

\begin{table*}
  \centering
  \caption{Log of the pre-perihelion observations. $\Delta_{tp}$
  indicates time since perihelion, negative before and positive after perihelion, $r_h$ is the heliocentric distance, 
$\Delta$ is the geocentric distance, \texttt{PsAng} is the position angle of the Sun to comet radius vector, \texttt{PlAng} is the angle between the Earth and the comet orbital plane, and \texttt{Phase} is the phase angle. The `Label' column provides a single-letter identifier for each image}. ($B_{max}$) indicates the innermost isophote levels in the images of Figure \ref{fig:contours}  ( labels a to e in column 2), in mag arcsec$^{-2}$, and the last column displays the projected dimensions on the sky plane at the comet's nucleus distance of those images.   
  \label{tab:logobs_pre}
  \begin{tabular}{|c|c|c|c|c|c|c|c|c|c|c|}
    \hline
    Instrument/Telescope &  Label & Time & $\Delta_{tp}$ & r$_h$ & $\Delta$   & \texttt{PsAng} & \texttt{PlAng} & \texttt{Phase} & $B_{max}$ & Dimensions \\
 & & (UT) & (days) & (au) &  (au) & ($^\circ$) & ($^\circ$) &($^\circ$) & (mag arcsec$^{-2}$) & (km$\times$km) \\
    \hline        
CAFOS/CAHA 2.2m  &a&2025-Jul-17 20:59& --103.6 &  3.95&  3.06&  100.8 & --0.74 & 8.02 & 20.5 & 23525$\times$23525\\
CAFOS/CAHA 2.2m  &b&2025-Jul-20 21:18& --100.6 &  3.86&  3.00&  100.3 & --0.68 & 9.25 & 20.5 & 23064$\times$23064\\
CAFOS/CAHA 2.2m  &c&2025-Aug-02 20:59& --87.6 &  3.43&  2.79&  100.3 & --0.37 & 14.6 & 19.0 & 21449$\times$21449\\
ALFOSC/NOT 2.5m &d& 2025-Aug-15 22:11 & --74.5 & 3.01 & 2.65 & 101.5 & 0.01 & 19.3 & 19.5 & 28791$\times$24678\\
ALFOSC/NOT 2.5m &e& 2025-Aug-28 21:18 & --61.6 & 2.61 & 2.59 & 102.7 & 0.42 & 22.4 & 19.0 & 40200$\times$32160 \\
\hline
\end{tabular}
\end{table*}

\begin{table*}
  \centering
  \caption{Log of the post-perihelion observations. All images were acquired with the FERVOR-L camera attached to the TST 1-m telescope at Teide Observatory. $\Delta_{tp}$
  indicates time range covered since perihelion, and $r_h$,  
$\Delta$, \texttt{PsAng}, \texttt{PlAng}, and \texttt{Phase}  are the ranges covered in heliocentric distance, geocentric distance, position angle of the Sun-to-comet radius vector, angle between the Earth and the comet orbital plane, 
and phase angle, respectively}.   
  \label{tab:logobs_post}
  \begin{tabular}{|c|c|c|c|c|c|c|c|c|}
    \hline
    Year & Month & Days & $\Delta_{tp}$(days) & r$_h$ (au)& $\Delta$ (au)  & \texttt{PsAng}($^\circ$) & \texttt{PlAng} ($^\circ$) & \texttt{Phase}($^\circ$) \\
    \hline        
2025 & 12& 2-5,7-9,11, &  33-62  & 1.82-2.63 & 1.80-1.90 & 291.1-293.6 & 1.79-2.39 & 16.5-30.6 \\ 
    &     & 18-20,23-26,28, &    &   &   &   &  &  \\
    &     & 30,31           &     &   &  &  &  & \\
2026 & 01 & 9-16, 19-30 & 71-92 & 2.92-3.59 & 2.00-2.62 & 90.19-290.7 & 0.39-1.70 & 0.74-14.9 \\
2026 & 02 & 7-20 & 100-113 & 3.85-4.28 & 2.94-3.50 & 93.8-94.3 & --0.21-0.11 & 6.4-9.3 \\

\hline
\end{tabular}
\end{table*}

\begin{table*}
  \centering
  \caption{Selected post-perihelion median-stack images for modelling. All quantities refer to the mean time of the observations. $\Delta_{tp}$
  indicate time after perihelion passage, and $r_h$,  
$\Delta$, \texttt{PsAng}, \texttt{PlAng}, and \texttt{Phase} are the mean heliocentric distance, geocentric distance, position angle of the Sun-to-comet radius vector, angle between the Earth and the comet orbital plane, 
and phase angle, respectively.  The `Label' column provides a single-letter identifier for each image}. ($B_{max}$) indicates the innermost isophote levels in the images of Figure \ref{fig:contours}  ( labels f to o in column 2), and the last column displays the projected dimensions on the sky plane at the comet's nucleus distance of those images.
  \label{tab:sample_dates_post}
  \begin{tabular}{|c|c|c|c|c|c||c|c|c|c|c|}
    \hline
Instrument/Telescope &  Label &  Time  & $\Delta_{tp}$ & r$_h$  &  $\Delta$   & \texttt{PsAng} & \texttt{PlAng} & \texttt{Phase}& $B_{max}$ & Dimensions\\
 & & (UT) & (days) & (au) & (au) & ($^\circ$) & ($^\circ$)&
 ($^\circ$) & (mag arcsec$^{-2}$) & (km$\times$km) \\
    \hline        
FERVOR-L/Teide TST-1m & f &2025-Dec-02 05:37 &  +33.7 &  1.83 &  1.90 & 293.6 &    2.40&   30.58 & 20.0 & 372060$\times$268710 \\
FERVOR-L/Teide TST-1m & g& 2025-Dec-07 05:15 &  +38.7 &  1.96 &  1.85 & 293.4 &   2.38 &  29.82 & 20.0  & 362295$\times$261676\\
FERVOR-L/Teide TST-1m &h &2025-Dec-18 04:27 & +49.7  & 2.26  & 1.80  & 292.5 &   2.20 &  24.96 & 20.0 & 352485$\times$254573\\
FERVOR-L/Teide TST-1m & i &2025-Dec-27 03:49 & +58.7  & 2.52  & 1.83  & 291.4 &   1.90 &  18.68 & 20.0  & 358380 $\times$ 258830\\
FERVOR-L/Teide TST-1m & j &2026-Jan-11 03:22 & +73.7  & 2.98  & 2.06  & 291.2 &   1.22 &  7.21 & 20.0 & 313740$\times$179280 \\
FERVOR-L/Teide TST-1m &k &2026-Jan-26 23:48 & +89.5  & 3.49  & 2.52  & 84.5  &  0.50  &  2.33 & 20.5 & 328980$\times$ 175456\\
FERVOR-L/Teide TST-1m &l &2026-Feb-04 22:18 & +98.4  & 3.78  & 2.86  & 93.4  &  0.18  &  5.77 & 20.5 & 186690$\times$124426\\
FERVOR-L/Teide TST-1m &m &2026-Feb-06 22:16 & +100.4 & 3.85  & 2.94  & 93.8  &  0.12  &  6.39  & 20.5  & 191910$\times$140734\\
FERVOR-L/Teide TST-1m & n&2026-Feb-09 22:24 & +103.4 & 3.95  & 3.06  & 94.1  &  0.03  &  7.23 & 20.5 & 186424$\times$106528\\
FERVOR-L/Teide TST-1m & o &2026-Feb-18 22:10 & +112.4 & 4.25  & 3.47  & 94.3  & -0.19  &  9.10 & 21.0 & 75500$\times$52850 
 \\
\hline
\end{tabular}
\end{table*}

\begin{figure*}
\includegraphics[angle=-90,trim=1cm 1cm 0cm 1cm,clip,width=19cm]{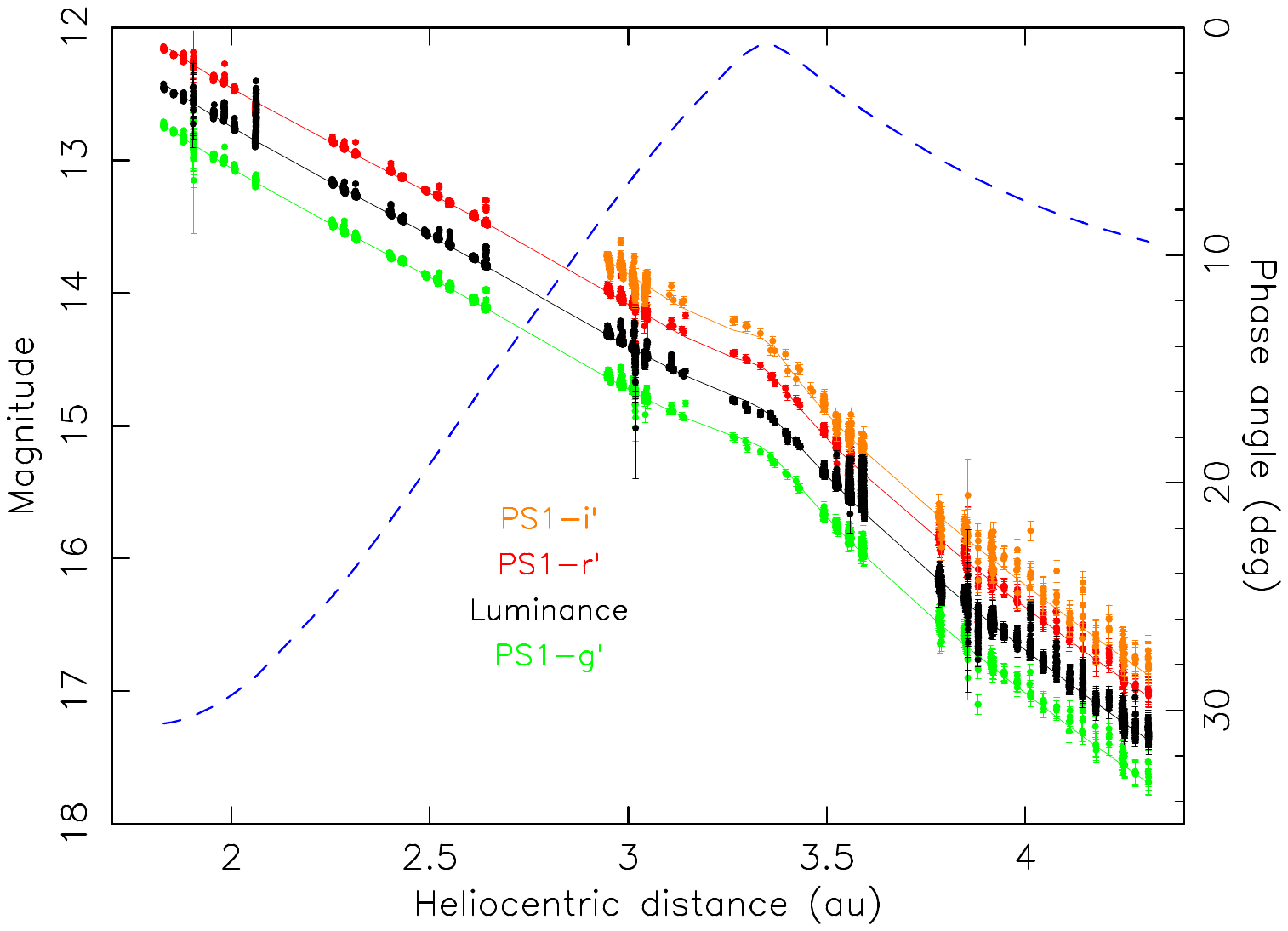}   
\caption{Magnitude (left vertical axis) vs heliocentric distance post-perihelion within circular apertures of radius 15,000 km, for each photometric band, as indicated by the different colours. The fits to the data (solid lines) were performed using Equation \ref{eq:photometry}. The blue dashed line shows the comet phase angle (right vertical axis) vs heliocentric distance. Note the conspicuous shoulders in each fitting curve near minimum phase angle caused by the backscattering enhancement of the dust particles.}
\label{fig:mag-vs-rh}
\end{figure*}

\begin{figure*}
\includegraphics[angle=-90,trim=1cm 1cm 0cm 1cm,clip,width=19cm]{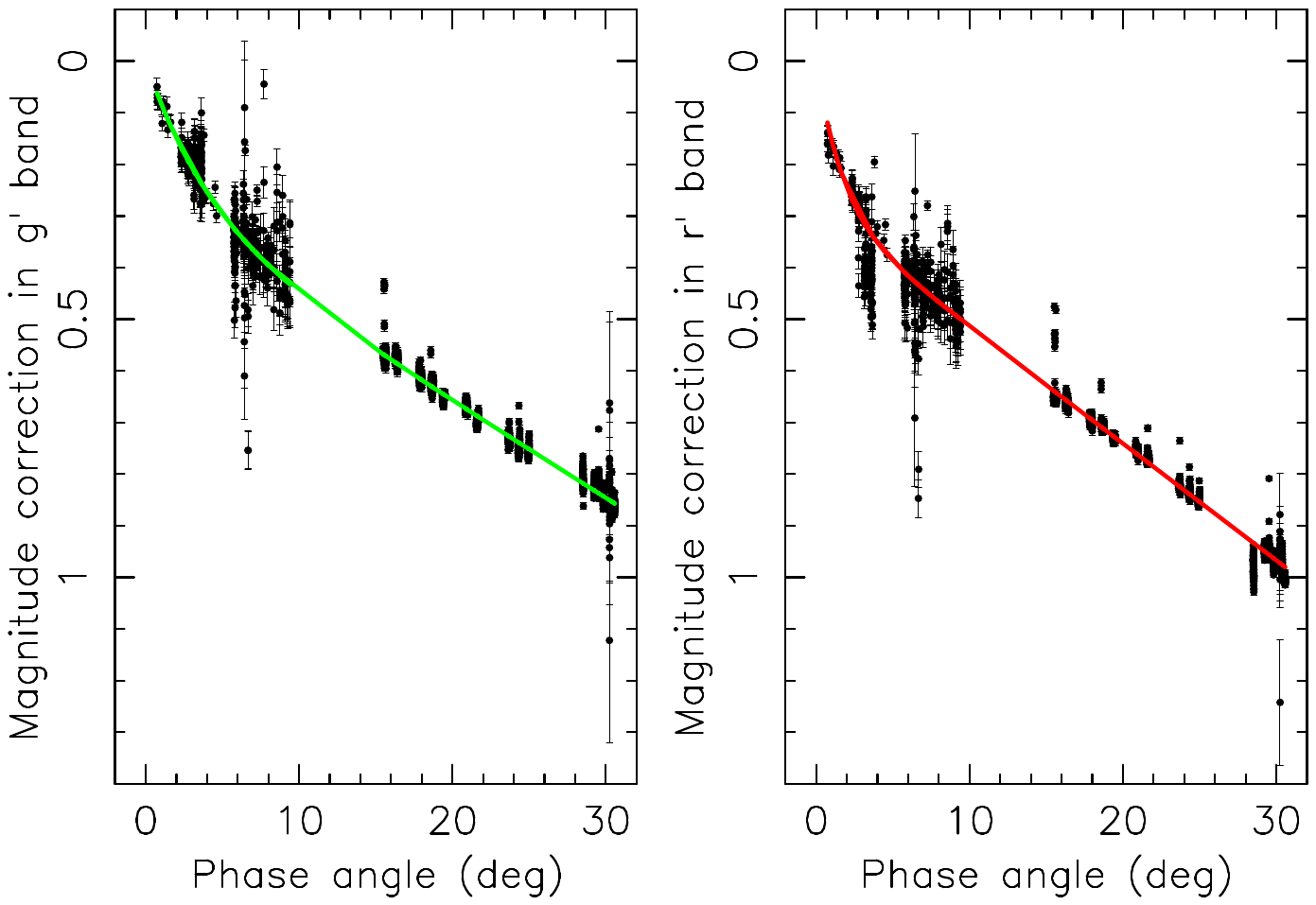}   
\caption{Magnitude correction for filters $g'$ (left panel) and $r'$ (right panel) as a function of phase angle from the post-perihelion images, highlighting the opposition surge near the backscattering direction in both wavelengths. The data points are indicated by solid circles, and the thick solid lines are the fits following Equation \ref{eq:photometry}.}
\label{fig:opposition-effect}
\end{figure*}

\section{Post-perihelion color and phase function of the dust}  \label{sec:phase_function}

For the post-perihelion images, we performed aperture photometry around the photocentre of each image to compute the magnitude as a function of time for each bandpass. To perform this task we used the routine \texttt{APER} in the \texttt{IDL} Astronomical Library with three aperture radii, namely 5,000, 10,000, and 15,000 km,  to estimate the total flux. The sky background was estimated from an annulus well outside the coma (inner radius $\sim$300,000 km, width $\sim$50,000 km), chosen to avoid any coma signal. The same sky annulus was used for the three apertures. The derived \textit{apparent} colour indices of the dust are essentially constant with the aperture used (see Table \ref{tab:colorindex}), and have mean values of  $g'-r'$=0.62$\pm$0.02 mag, and $r'-i'$=0.18$\pm$0.05 mag. Using these colour indices and the transformation equations between PS1 and SDSS photometric systems  \citep{2016ApJ...822...66F}, we obtain $g-r$=0.70$\pm$0.02 and $r-i$=0.22$\pm$0.05, that are close to the range of the values found for most comets in the Solar System as reported by \cite{2012Icar..218..571S} ($g-r$=0.62$\pm$0.05 mag and $r-i$=0.23$\pm$0.05 mag). No significant variations in those color indexes as a function of the heliocentric distance are found (see Figure \ref{fig:mag-vs-rh} below). Also, for the other interstellar comet known to exhibit activity, 2I/Borisov, \citet{2019ApJ...886L..29J} reported colour indices of $B-V = 0.80 \pm 0.05$, $V-R = 0.47 \pm 0.03$, and $R-I = 0.49 \pm 0.05$. Using the SDSS to Johnson-Cousins photometric transformations of \citet[][their Table 3]{2006A&A...460..339J}, 
specifically  $g-r = 1.646(V-R)-0.139$ and $r-i=1.007(R-I)-0.236$ 
 our derived $g-r$ and $r-i$ colour indices correspond to $V-R \approx 0.51$ and $R-I \approx 0.45$, values that are close to those obtained for interstellar comet~2I/Borisov.


Recent high-resolution observations have provided new insight into the phase behaviour of 3I/ATLAS near and after perihelion. In particular, \texttt{HST} imaging obtained at epochs close to ours has revealed unexpected features in the phase function, including evidence for an opposition surge. From the analysis of these post-perihelion data, \citet{2026arXiv260121569H} reported a linear phase--function coefficient of $\beta_{\alpha} = 0.026 \pm 0.006$ in the broadband F350LP filter and identified an opposition effect not previously detected in any comet. Following their approach, we performed, for each band, a photometric fit to the apparent magnitude measurements using the relation:

\begin{equation}
m = H + 2.5n\log_{10}r_h + 5 \log_{10}\Delta + \beta_\alpha\alpha + m_{oe}
\{\,1 - \exp(-\alpha/w_{oe})\,\}
\label{eq:photometry}
\end{equation}

In this expression, $H$ is the absolute magnitude, $n$ is the activity index, $m_{oe}$ is the magnitude of the opposition effect, and $w_{oe}$ is the e-folding width of the opposition effect. To perform the fits we used the downhill simplex minimization technique described by \cite{amoeba} and implemented in the Numerical Recipes book, routine \texttt{AMOEBA} \citep{1986nras.book.....P}.

\begin{table}
\centering
\caption{Computed \textit{apparent} colour indices and uncertainties, as a function of the aperture radius, together with the averaged values.}
\label{tab:colorindex}
\begin{tabular}{ccc}
\hline
Aperture & $g'-r'$ & $r'-i'$  \\
(km) & (mag) & (mag)   \\
\hline
5,000 & 0.61$\pm$0.03 & 0.17$\pm$0.12 \\
10,000& 0.63$\pm$0.03 & 0.19$\pm$0.06 \\
15,000 & 0.62$\pm$0.04 & 0.19$\pm$0.04 \\
\hline
Mean & 0.62$\pm$0.02 & 0.18$\pm$0.05   \\
\hline
\end{tabular}
\end{table}
 
 \begin{table*}  
\caption{The resulting values of the fitting parameters in Equation \ref{eq:photometry} as a function of the aperture radius and photometric band.}
\centering
\label{tab:photometry}
\begin{tabular}{lcccc}
\toprule
 & \multicolumn{4}{c}{Ap=5,000 km} \\ 
\midrule
 & $g'$ & $r'$ & $i'$ & $Lum$ \\
\midrule
$H$ (mag) &8.3$\pm$0.1 & 7.59$\pm$0.09 & 6.8$\pm$0.2 & 7.85$\pm$0.01 \\
$\beta_\alpha$ (mag deg$^{-1}$)&0.034$\pm$0.001 & 0.037$\pm$0.001 & 0.02$\pm$0.01 & 0.033$\pm$0.001 \\
$n$ &4.91$\pm$0.03 & 5.01$\pm$0.04 & 5.27$\pm$0.05 & 4.94$\pm$0.01 \\
$m_{oe}$ (mag) &0.2$\pm$0.1 & 0.1$\pm$0.1 & 0.4$\pm$0.1 & 0.25$\pm$0.01 \\
$w_{oe}$ (deg)& 3$\pm$1 & 1.0$\pm$0.5 & 2$\pm$1 & 3.2$\pm$0.1 \\
\midrule
& \multicolumn{4}{c}{Ap=10,000 km} \\
\midrule
 & $g'$ & $r'$ & $i'$ & $Lum$ \\
 \midrule
$H$(mag)&7.87$\pm$0.04 & 7.18$\pm$0.07 & 6.6$\pm$0.1 & 7.54$\pm$0.02 \\
$\beta_\alpha$ (mag deg$^{-1}$)& 0.023$\pm$0.001 & 0.026$\pm$0.001 & 0.018$\pm$0.002 & 0.024$\pm$0.001 \\
$n$ & 4.46$\pm$0.02 & 4.50$\pm$0.02 & 4.72$\pm$0.03 & 4.49$\pm$0.01 \\
$m_{oe}$ (mag) &  0.31$\pm$0.03 & 0.27$\pm$0.06 & 0.4$\pm$0.1 & 0.29$\pm$0.02 \\
$w_{oe}$ (deg)&  2.7$\pm$0.3 & 1.4$\pm$0.3 & 1.6$\pm$0.9 & 3.2$\pm$0.2 \\
\midrule
& \multicolumn{4}{c}{Ap=15,000 km} \\
\midrule
 & $g'$ & $r'$ & $i'$ & $Lum$ \\
 \midrule
$H$(mag) &7.5$\pm$0.1 & 6.89$\pm$0.04 & 6.4$\pm$0.2 & 7.26$\pm$0.01 \\
$\beta_\alpha$ (mag deg$^{-1}$) & 0.021$\pm$0.001 & 0.024$\pm$0.001 & 0.018$\pm$0.002 & 0.021$\pm$0.001 \\
$n$ & 4.35$\pm$0.02 & 4.34$\pm$0.01 & 4.55$\pm$0.03 & 4.34$\pm$0.01 \\
$m_{oe}$ (mag)& 0.3$\pm$0.1 & 0.28$\pm$0.04 & 0.4$\pm$0.2 & 0.3$\pm$0.1 \\
$w_{oe}$ (deg)& 3$\pm$1 & 1.5$\pm$0.2 & 1.5$\pm$0.4 & 3.4$\pm$0.1  \\

\bottomrule
\end{tabular}
\end{table*}

The numerical values of the fitting parameters for the three aperture sizes already mentioned are displayed in Table \ref{tab:photometry}. The resulting fits for aperture $Ap$=15,000 km (for which the best signal-to-noise measurements of magnitude and smallest errors are found)  are shown in Figure \ref{fig:mag-vs-rh}.  The uncertainties in the best‑fit parameters were estimated by Monte Carlo resampling of the photometric data. For each trial, synthetic magnitudes were generated by adding Gaussian noise consistent with the 2-$\sigma$ uncertainties returned by the routine \texttt{APER}, and the model was refitted using \texttt{AMOEBA}.  The 2-$\sigma$ dispersion of the resulting distribution of the best‑fit parameters was adopted as the parameter uncertainty. 

The activity index $n$ is found to decrease slightly, but significantly, with aperture size, for all photometric bands, with values consistent with that reported by \cite{2026arXiv260121569H} of $n$=4.7$\pm$0.2 for the $V$ band post-perihelion. We also found a slight but significant increase in $n$ with wavelength from the $r'$ to the $i'$ 
band for apertures of 5,000 and 10,000~km, whereas no such clear trend is present for the 
15,000~km aperture. Regarding the magnitude of the opposition effect, most values are located between 0.2 and 0.4 mag, in agreement with that found by 
 \cite{2026arXiv260121569H}.  The e-folding width values range from $\sim$ 1 to 3 deg, again in line with that reported by  \cite{2026arXiv260121569H}, and confirming the presence of a significant opposition effect. To highlight this effect, in Figure \ref{fig:opposition-effect} we plot the $r$ band magnitude correction vs. phase angle, which clearly shows the opposition surge close to 0 deg phase angle. 
 
 On the other hand, the linear phase function coefficient shows a significant decrease with aperture size for all photometric bands. For instance, in the $r$ filter it varies from $\beta_{\alpha}$=0.037$\pm$0.001 mag deg$^{-1}$ for Ap=5,000 km to $\beta_{\alpha}$=0.024$\pm$0.001 mag deg$^{-1}$ for Ap=15,000 km. This can be interpreted as the result that for larger apertures the dust population becomes dominated by smaller grains that scatter light more isotropically, reducing the phase slope.

\section{Monte Carlo dust tail modelling}  \label{sec:Model}

To derive the physical properties of the dust and the corresponding dust mass loss rate, we used our forward Monte Carlo dust‑tail code. A complete description of the numerical model is given in 
\cite{2025A&A...695A.263M}, so only a summary is provided here. The code is applicable at distances larger than about 20 nucleus radii, where both the nucleus gravity and the gas‑drag force become negligible. Under these conditions, the initial particle velocities correspond to the terminal speeds reached at those distances. The dust particles,  treated as compact spheres, are assumed to follow Keplerian orbits under the sole action of solar gravity and radiation pressure.

The particle trajectories depend on their initial velocity and on the ratio of radiation pressure to solar gravity, expressed by the parameter $\beta$ \citep[see][]{1968ApJ...154..327F} defined as:
\begin{equation}
\beta = \frac{C_{pr} Q_{pr}}{2 \rho{\rm_d} r},
\label{eq:beta}
\end{equation}
where \(C_{pr} = 1.19 \times 10^{-3}\,\mathrm{kg\,m^{-2}}\) is the radiation‑pressure coefficient, \(Q_{pr}\) is the radiation‑pressure efficiency (taken as unity for absorbing particles larger than \(\sim 1\,\mu\mathrm{m}\) \citep{1979Icar...40....1B}), \(\rho{\rm_d} = 800\,\mathrm{kg\,m^{-3}}\) is the particle density and \(r\) is the particle radius. The particle density is taken from that derived by 
\citep{2015ApJ...802L..12F} from the \texttt{GIADA} instrument measurements on board \texttt{Rosetta} mission.  For each particle, its heliocentric position at the observation epoch is computed and projected onto the sky plane. In the Monte Carlo simulation, we generate a large number of ejected particles, typically $\gtrsim 10^{7}$. Their brightness depends on particle size, geometric albedo,  and phase angle of observation. Typical cometary dust is very dark, with geometric albedos of 
$p_{v0} \sim 0.04$ \citep{1981Icar...47..342H,2004come.book..577K}, a value that we adopted in our simulations. Phase‑angle effects can be  corrected using the empirical dust‑phase function derived by David Schleicher\footnote{\url{https://asteroid.lowell.edu/comet/dustphase/details}}, or the recently derived by \cite{2025P&SS..26506164B}, who fitted a two-term Henyey-Greenstein phase function to a large compilation of published short- and long-period comet phase-function measurements, which can be considered representative of the average behavior of most comets in the Solar System. However, in this case we used the linear-exponential phase function $\phi(\alpha)$ resulting from the fitting of our post-perihelion images in the $r$-SDSS filter, so we can write:

\begin{equation}
  \phi(\alpha)=  \beta_\alpha \alpha + m_{oe}
\{\,1 - \exp(-\alpha/w_{oe})\,\} 
\label{eq:linear-exponential}
\end{equation}
where $\beta_\alpha$=0.024 mag deg$^{-1}$, $m_{oe}$=0.28 mag, and $w_{oe}$= 1.5 deg are the corresponding fitting parameters (see Table \ref{tab:photometry}).  

\begin{figure}
    \centering
    \includegraphics[angle=-90,trim=4cm 3cm 2cm 4cm,clip,width=\linewidth]{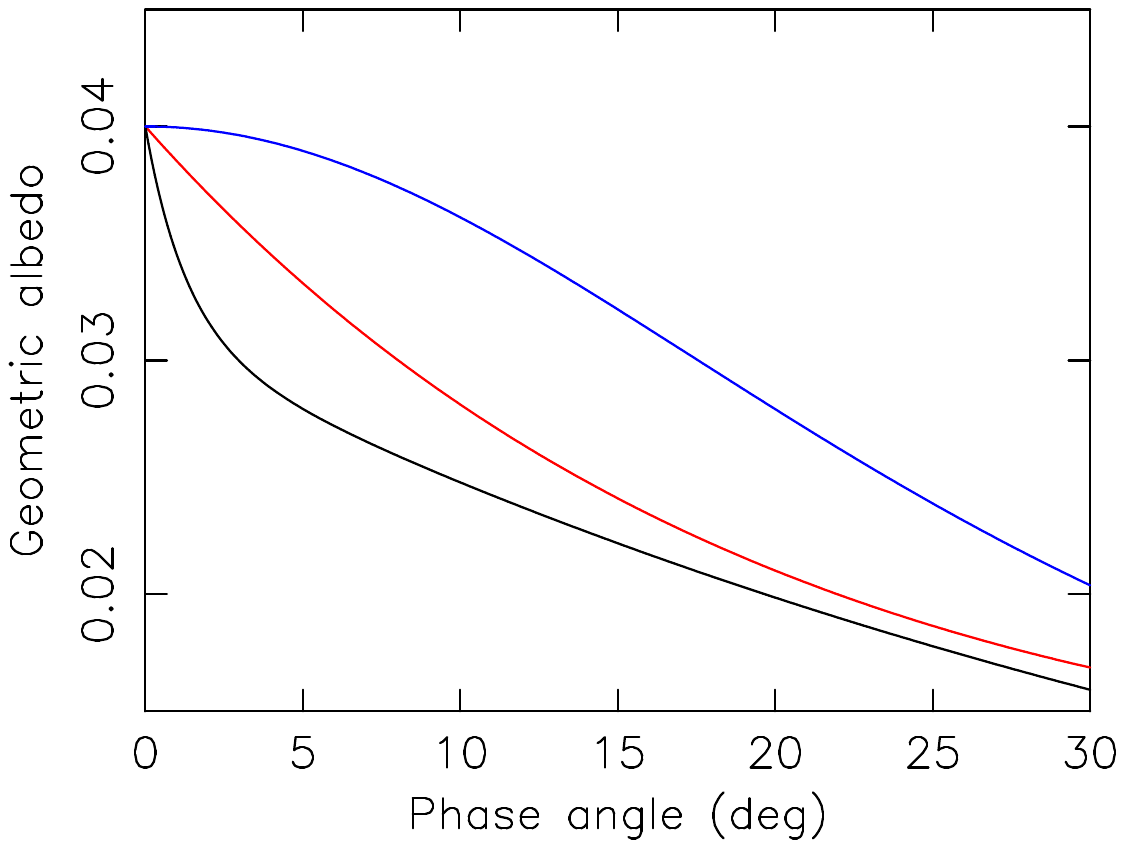}
\caption{Geometric albedo as a function of phase angle for the linear--exponential fit to the $r'$ filter magnitude measurements using the 15{,}000~km aperture radius (black line), compared with the phase curve derived by Schleicher (red line) and that by Bertini et al. (2025) (blue line).}
    \label{fig:geometric-albedo}
\end{figure}

The total backscattering enhancement (\texttt{BSE}), of the linear-exponential model, defined as the ratio of the geometric albedo at phase angles of $0^{\circ}$ and $30^{\circ}$, is \texttt{BSE}$\sim$2.5, which lies within the interval measured for most Solar System comets \citep[\texttt{BSE}=2--3.5, see][]{2019MNRAS.482.2924B}, and close to the upper limit of laboratory scattering measurements for comet dust analogues \citep[\texttt{BSE}=1.9--2.3, see][]{2019MNRAS.484.2198F}. Figure~\ref{fig:geometric-albedo} displays the linear–exponential geometric albedo model together with those derived by David Schleicher and by \citet{2025P&SS..26506164B}, where all curves are normalised to $p_{v0}=0.04$. The Schleicher curve shows a \texttt{BSE} comparable to the linear–exponential model, whereas the Bertini et al. curve is somewhat flatter (\texttt{BSE}$\sim$2). For the interstellar comet 2I/Borisov, \citet{2019ApJ...886L..29J} adopted a linear phase coefficient of 0.04 mag\,deg${^{-1}}$ to correct their measurements for phase-angle effects, which yields a flux ratio of 1.7 between $0^{\circ}$ and $15^{\circ}$, consistent with the value extracted from our linear–exponential model (1.8). In contrast to the smoother behaviour displayed by most Solar System comets, the linear–exponential curve for 3I/ATLAS exhibits a pronounced opposition surge—a very steep increase in brightness between phase angles $\approx5^{\circ}$ and $0^{\circ}$—that has not been previously observed in any comet. Such behaviour arises naturally from highly porous aggregate particles, whose internal multiple scattering and coherent backscattering effects generate a stronger opposition surge than in compact grains of the same size \citep[e.g.][]{2011JQSRT.112.2193M,2006JQSRT.100..199K,2006JQSRT.100..220L,2004come.book..577K}. In this context, the behaviour of the Schleicher and Bertini et al. phase functions could suggest that the dust grains released by 3I/ATLAS possess a higher porosity than the average particles inferred for typical Solar System comets. A definitive demonstration of this, however, lies well beyond the scope of the present work.

The particle terminal velocities are parameterized as a function of size, heliocentric distance, and solar zenith angle \(z\) at the emission point:

\begin{equation}
v(\beta,t,z) = v_0\,\beta^\gamma\,v_1(t)\,(\cos z)^\varepsilon.
\label{eq:speed}
\end{equation}

Here, \(t\) is time, \(v_0\) is a constant, and the exponent \(\gamma\) is usually restricted to \(\gamma \lesssim 0.5\) \citep{1987A&A...171..327F}. In our simulations we take $\gamma$=0.5. The speed factor \(v_1(t)\) is frequently expressed as \(v_1(t) = r_h^\Gamma\), where \(r_h\) is the heliocentric distance and \(\Gamma = -0.5\) \citep{1951ApJ...113..464W,2000Icar..148...80R}. The dependence on the solar zenith angle is a new feature of the current version of the code. It breaks the usual one‑to‑one correspondence between particle size and terminal velocity found in earlier dust‑tail models \citep[see][]{2011ApJ...732..104T}, since particles of the same size may reach different terminal speeds depending on \(z\). As discussed in 
\cite{2025MNRAS.539..949M}, this dependence is supported by 3‑D dusty‑coma models \citep{1997Icar..127..319C,2011ApJ...732..104T}.  Following \cite{2008Icar..193..572K} and \cite{2009AJ....137.4633K}, the exponent \(\varepsilon\) may take values of 1 or 0.25 depending on whether the velocity scales with insolation or with the surface temperature of a slowly rotating nucleus. In our case, we adopted $\varepsilon = 1$, based on the conspicuous sunward–pointing tail morphology visible in the early HST images \citep{2025ApJ...990L...2J}. 

The nucleus contributes to the central pixel brightness of the images, although this effect is only noticeable far from perihelion when the activity is weak. For the nucleus we assume the same geometric albedo as for the dust particles, corrected for phase angle using a linear phase coefficient of \(0.03\,\mathrm{mag\,deg^{-1}}\). The nucleus radius is set to \(R_N = 1\,\mathrm{km}\), in line with that reported by 
\cite{2026arXiv260121569H} from HST images (R$_N$=1.3$\pm$0.2 km), and with that inferred by non-gravitational forces  \citep{2026arXiv260315735T} (R$_N$=0.74-1.15 km).
Synthetic images are produced by summing the contribution of all simulated particles. The resulting brightness distribution depends on the dust production rate, particle velocities, and the assumed size distribution. The latter is described by a power law between minimum and maximum radii, \(r_{\min}\) and \(r_{\max}\), with exponent \(\kappa\), such that $n(r) \propto r^\kappa$.

 The dust tail model requires specifying the activity onset time and the dust production rate profile. As noted in Section \ref{sec:Introduction}, as the first observational indication of cometary activity occurs at about 6.4 au \citep{2025ApJ...991L...2F}, i.e., about 180 days before perihelion, the activity must necessarily have begun earlier. In our modelling, we therefore adopt an onset time 300 days before perihelion (at $r_h \approx 10.5$ au) as a physically plausible and conservative assumption. At that epoch we set the dust production rate to zero, ensuring a smooth transition from an inactive nucleus to the onset of measurable activity. Although marginal activity driven by the sublimation of CO or CO$_2$ could in principle have occurred at even larger heliocentric distances in pristine nuclei \citep[e.g.,][]{2004come.book..317M}, the model predictions, both the modelled magnitudes and the synthetic brightness images, are completely insensitive to low activity levels at very far heliocentric distances. The dust production rate at the early epochs after the onset time onwards will be adjusted so that the measured TESS magnitudes  \citep{2025ApJ...991L...2F} are well reproduced by the model, as will be shown in Section \ref{sec:results}.

To perform the model simulations, we adopted the simplest possible scenario, assuming a constant size distribution along the orbit. In practice, this means that the minimum and maximum particle radii, as well as the exponent of the power-law size distribution, were kept fixed at all heliocentric distances.  This choice is consistent with the observational evidence that the color indices of the comet remained essentially constant with heliocentric distance, at least along the post-perihelion branch, suggesting that no significant evolution in the dust size distribution occurred during this phase. We likewise assumed a constant value of the parameter $v_0$ in the ejection-velocity law (Equation~\ref{eq:speed}).

The pre-perihelion visible polarimetric measurements by \cite{2025ApJ...992L..29G} show a deep negative branch and a low inversion angle, indicating the presence of large particles composed of a mixture of ice and dark refractory material. Post-perihelion polarimetric measurements by \cite{2026arXiv260108591C} exhibit consistent behavior. Guided by these findings, we adopted a minimum particle radius of $r_{\min} = 10\,\mu\mathrm{m}$. For different choices of maximum particle size, power-law size-distribution indices, and the parameter $v_0$, we varied the dust loss rate to obtain simultaneous fits to all images and to the evolution of the observed magnitudes.  In addition to the TESS magnitude data cited above, the pre-perihelion magnitude data are taken from the amateur association \texttt{Cometas\_Obs}, while for the post-perihelion branch we use both the \texttt{Cometas\_Obs} measurements and our own $r$-band data, which agree very well with the amateur observers estimates (see Figures~\ref{fig:lightcurve1} and \ref{fig:lightcurve}).  Although a fixed cometocentric projected aperture would have been more meaningful, the \texttt{Cometas\_Obs} magnitudes were obtained using a fixed square aperture of $10\arcsec \times 10\arcsec$ on the sky, thus sampling different cometocentric radii at different geocentric distances. To enable a self-consistent comparison, we computed the synthetic model magnitudes using a circular aperture with the same projected area of $100\arcsec^{2}$, corresponding to a radius of $\rho = \sqrt{100/\pi} = 5.64\arcsec$. We note that these magnitude measurements were not used as primary constraints on the dust physical properties, but as supplementary lightcurve data to extend the temporal baseline. All dust model parameters were determined from the isophote fits, while the light curve served solely to constrain the overall normalisation and temporal shape of the dust production rate profile.

\section{RESULTS AND DISCUSSION}
\label{sec:results}
\begin{figure*}
    \centering
    \includegraphics[angle=-90,trim=1cm 0cm 0cm 2.5cm,clip,width=\linewidth]{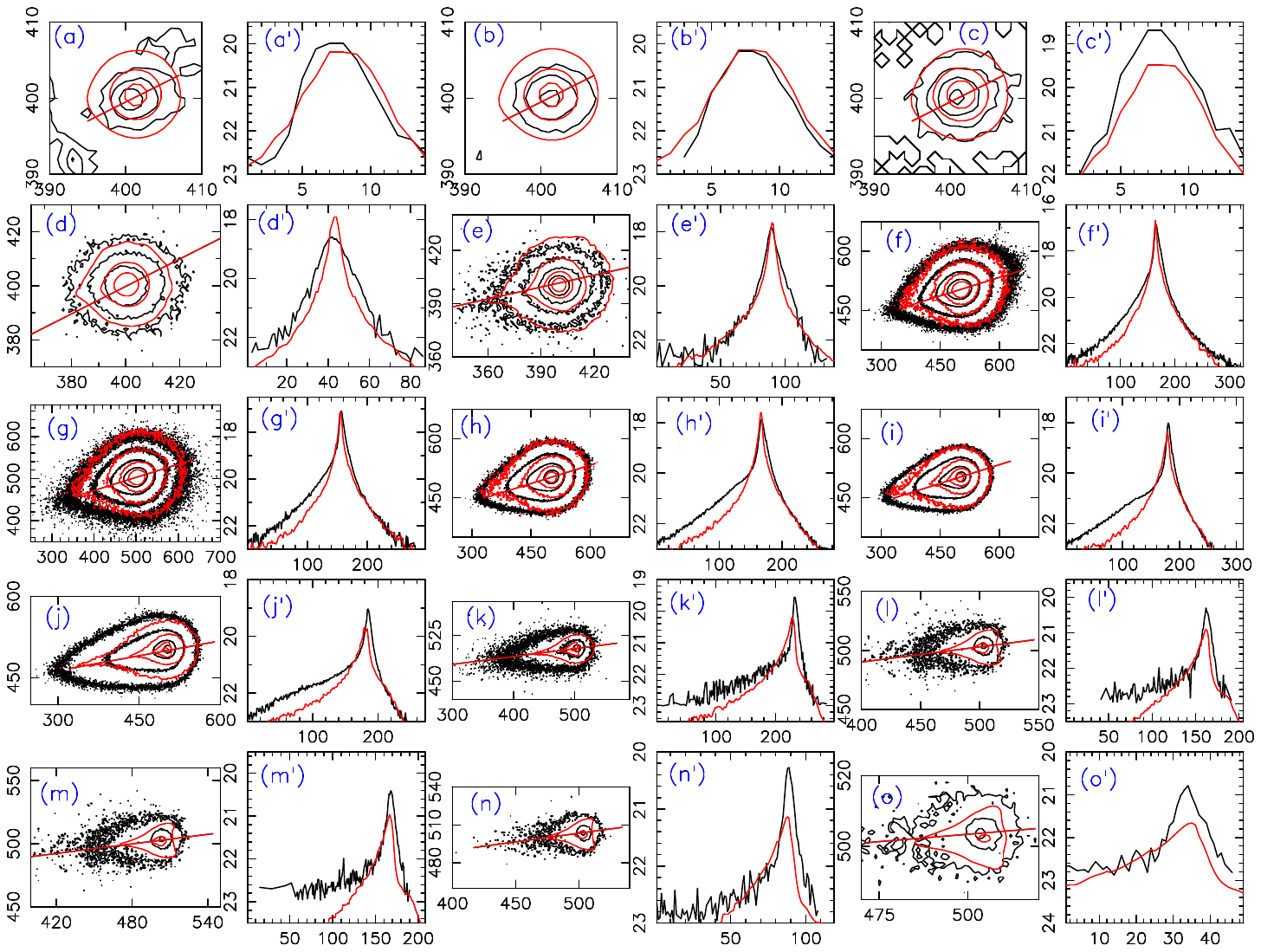}
\caption{Panels (a)–(o) show contour maps of the observed isophotes (black lines) together with the corresponding modelled isophotes (red lines) for the case $r_{\max}=1$~cm. For each of these contour panels, a companion plot is displayed to the right, labelled (a$'$)–(o$'$), showing a one-dimensional scan extracted along the direction indicated by the straight red line in the corresponding contour map. In these primed panels, the black and red curves represent the observed and modeled brightness profiles, respectively. The $x$- and $y$-axes in the contour plots are given in pixel units, and the innermost isophote levels as well as the physical dimensions of each image are listed in Tables~\ref{tab:logobs_pre} and \ref{tab:sample_dates_post}. In panels (a)–(o), isophote levels increase outward in steps of 1~mag. In the primed panels, the $x$-axis is expressed in pixel units and the $y$-axis in surface brightness units (mag~arcsec$^{-2}$).}
    \label{fig:contours1}
\end{figure*}

\begin{figure*}
    \centering
    \includegraphics[angle=-90,trim=1cm 0cm 0cm 2.5cm,clip,width=\linewidth]{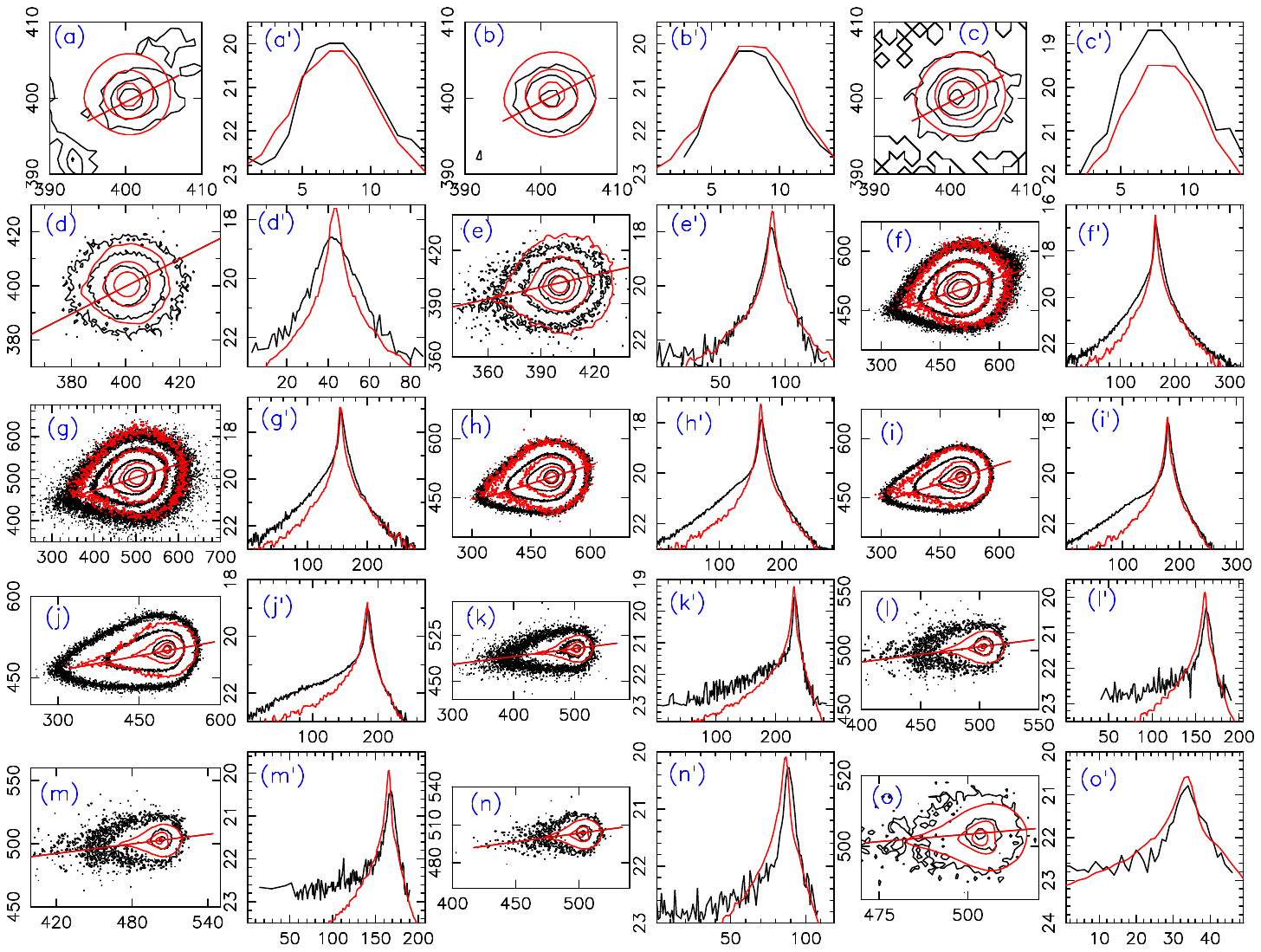}
\caption{As Figure \ref{fig:contours1}, but for $r_{max}=10$ cm.}
    \label{fig:contours}
\end{figure*}

\begin{figure}
    \centering
    \includegraphics[angle=-90,trim=3cm 1cm 0cm 3cm,clip,width=\linewidth]{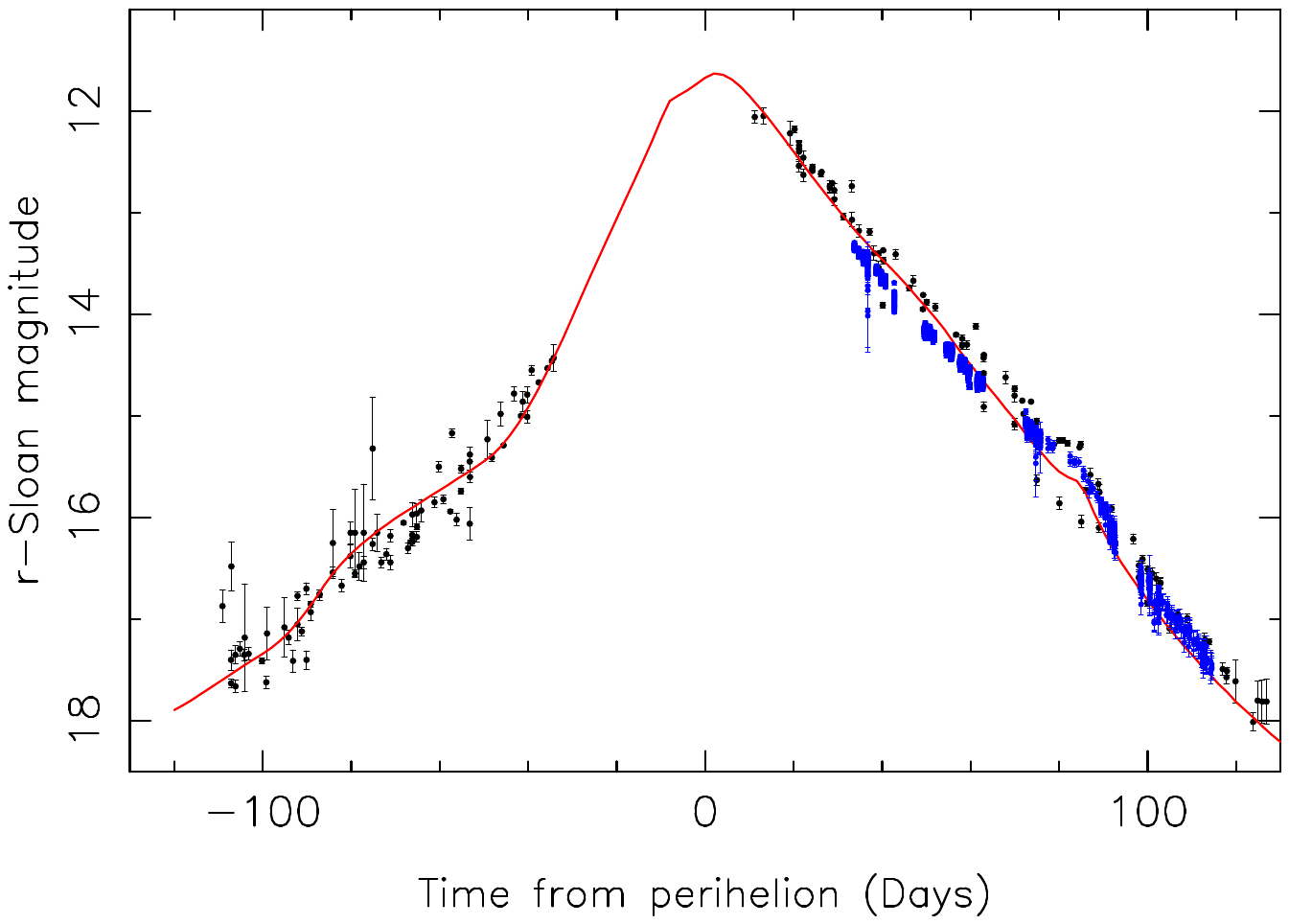}
\caption{Apparent magnitude in the $r$-SDSS filter vs. time from perihelion. The black dots are provided by the amateur association \texttt{Cometas\_Obs} for an aperture of 10\arcsec$\times$10\arcsec, the blue dots are those calculated for an equivalent-area circular aperture radius of 5.64\arcsec from the post-perihelion TST measurements, and the red line is the model magnitude calculations for the same aperture. This model has a maximum particle radius of $r_{max}$=1 cm}
    \label{fig:lightcurve1}
\end{figure}

\begin{figure}
    \centering
    \includegraphics[angle=-90,trim=3cm 1cm 1cm 3cm,clip,width=\linewidth]{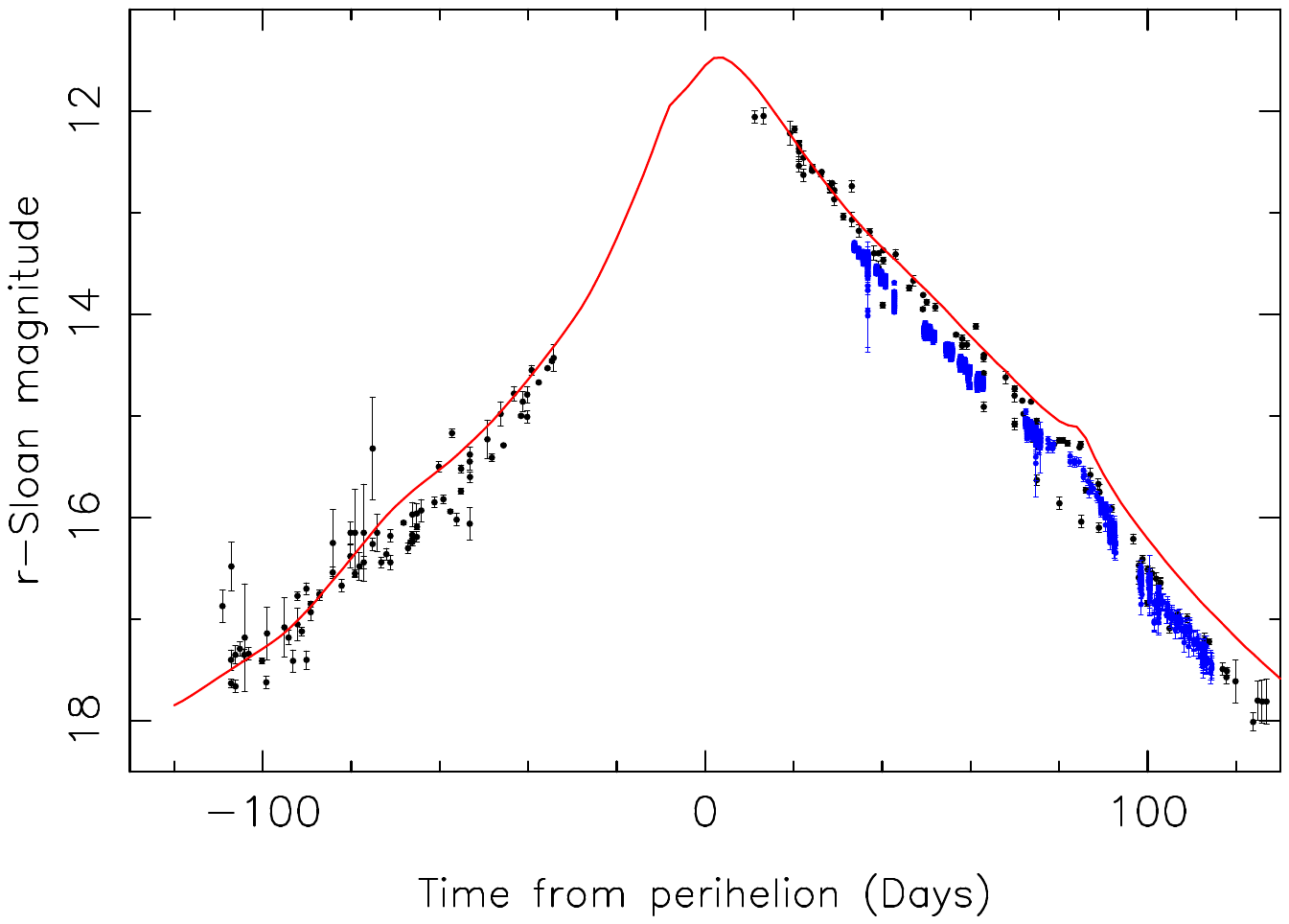}
\caption{As Figure \ref{fig:lightcurve1} but for a maximum particle radius of $r_{max}$=10 cm.}
    \label{fig:lightcurve}
\end{figure}

\begin{figure*}
    \centering
    \includegraphics[angle=-90,trim=1cm 0cm 2cm 2cm,clip,width=\linewidth]{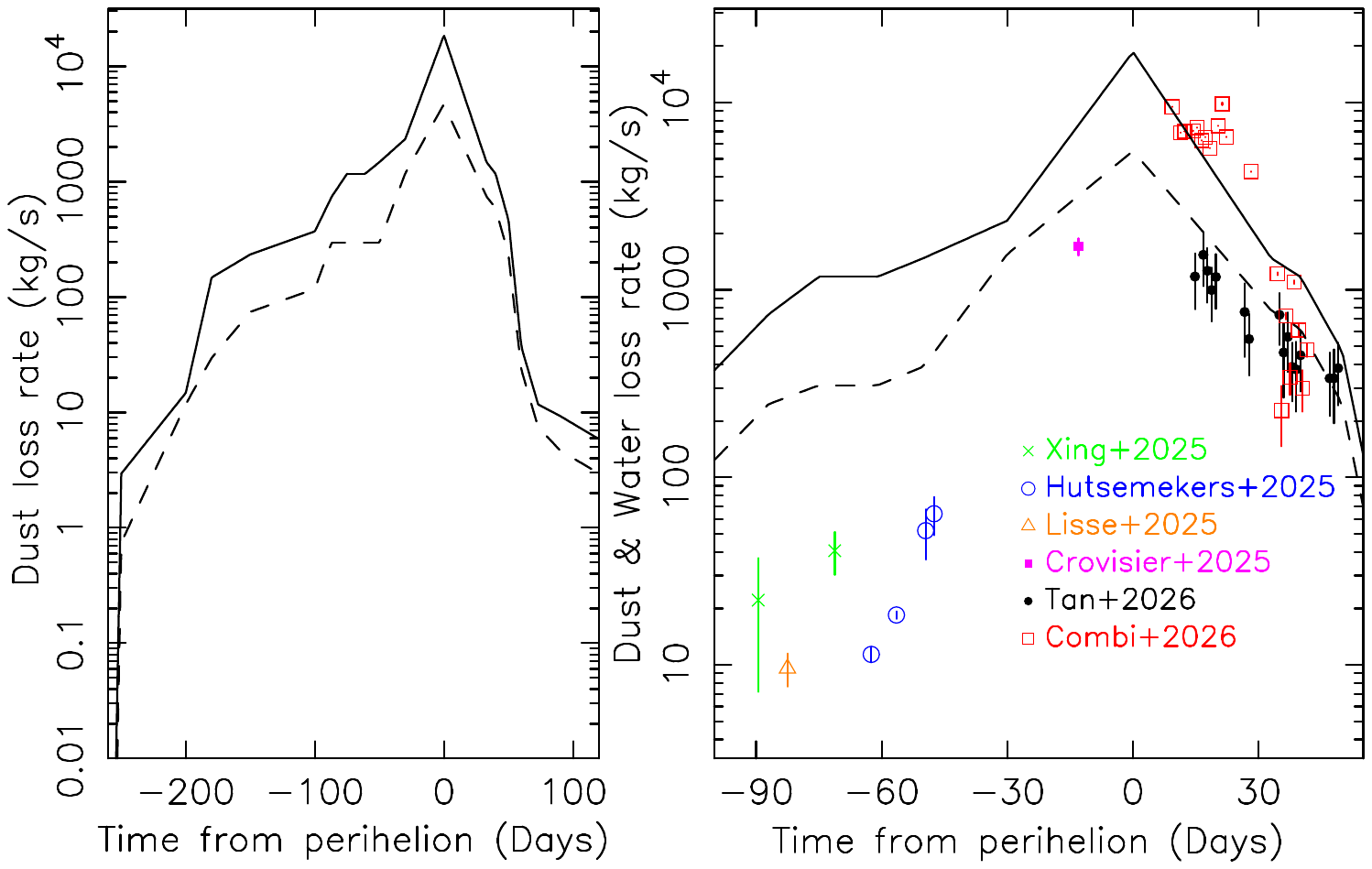}
\caption{Left panel: Derived dust mass loss rate as a function of heliocentric distance, assuming a dust outflow velocity parameter $v_{0} = 500~\mathrm{m\,s^{-1}}$ and a differential size distribution with slope $\kappa = -3.5$, minimum radius $r_{\min} = 10~\mu\mathrm{m}$, and two values of the maximum radius: $r_{\max} = 10~\mathrm{cm}$ (solid line) and $r_{\max} = 1~\mathrm{cm}$ (dashed line).
Right panel: Same dust mass-loss rate shown over a narrower time interval, together with published measurements of the water production rate by various authors, as indicated by the different colored symbols.}
    \label{fig:dustlossrate}
\end{figure*}

 The model depends on several physical parameters, as described above, that must be constrained from the observations; however, because the parameter space is large and each model evaluation is computationally expensive, a formal chi–square optimisation is not feasible. Instead, the parameters were adjusted iteratively through a structured trial–and–error exploration of the parameter space. In practice, one parameter was held fixed while the others were varied systematically within physically reasonable ranges, progressively refining the search until a reasonable and self‑consistent agreement with both the image isophotes and the magnitude lightcurve data was obtained. Following this procedure, 
 we found acceptable fits when  $v_{0}=500~\mathrm{m\,s^{-1}}$, with $r_{\min}=10~\mu\mathrm{m}$ as mentioned above, and a size--distribution slope of $\kappa=-3.5$, a value well within the range derived for many comets \citep[see, e.g.][]{2024come.book..653A}. This set of parameters --- $v_{0}=500~\mathrm{m\,s^{-1}}$, $\varepsilon=1$, $\kappa=-3.5$, and $r_{\min}=10~\mu$m --- constitutes what we refer to as the nominal model parameters. As for the maximum particle radius, we found that acceptable fits can be obtained for $r_{\max}$ values in the range 1--10~cm. 
 The contour plots of the modeled isophotes compared to the observations are shown in Figures~\ref{fig:contours1} and \ref{fig:contours}, while the corresponding fits to the lightcurve are presented in Figures~\ref{fig:lightcurve1} and \ref{fig:lightcurve}, for maximum particle radii of $r_{\max}=1$~cm and $r_{\max}=10$~cm, respectively.  As stated in the previous Section, the dust production rate at the earliest epoch after the onset time assumed (300 days pre-perihelion, $r_h\sim10.5$ au) was constrained with the magnitude data provided by the TESS observations \citep{2025ApJ...991L...2F}. These authors obtained TESS magnitudes at two epochs, 6.35--5.99 au and 5.92--5.47 au pre-perihelion, and converted them into $V$ magnitudes as listed in Table \ref{tab:TESS}. To further convert these $V$ magnitudes into r-SDSS magnitudes, we used our derived colour index  
 $g-r$ together with the photometric relations $g-r = 1.646(V-R)-0.139$ and $r-R=0.267(V-R)+0.088$ (valid for $(V-R)\leq0.93$) \citep[][, their Table 3]{2006A&A...460..339J}, obtaining $r=21.34\pm0.35$ and $r=19.79\pm0.35$ at the two epochs, respectively (see Table \ref{tab:TESS}). To match the observed magnitudes at those two epochs, together with the lightcurve data 
 and the observed image isophotes, the dust production rates have to follow the curves shown in Figure \ref{fig:dustlossrate}.  For the two maximum particle radii of the model, we obtain, at the epochs of TESS observations, the $r$ magnitudes show in Table \ref{tab:TESS}, that are in remarkable agreement with those observations.
 \begin{table}  
\caption{TESS magnitudes \citep{2025ApJ...991L...2F} compared with model magnitudes}
\centering
\label{tab:TESS}
\begin{tabular}{ccc|cc}
\hline
$r_h$   & TESS   & TESS  & Model   &Model  \\
range (au) & V-mag & $r$-SDSS & $r$-SDSS & $r$-SDSS  \\
            &      &         &$r_{\rm max}$=1 cm &$r_{\rm max}$=10 cm \\
 \hline
6.35--5.99  & 21.63$\pm$0.35 & 21.34$\pm$0.35 & 21.3--20.7 & 21.3--20.4 \\
5.92--5.47  & 20.08$\pm$0.35 & 19.79$\pm$0.35 & 20.4--19.6 & 20.2--19.4 \\
\hline
\end{tabular}
\end{table}
   
For each observed and modelled image, in addition to the contour plots, we also provide brightness scans along the red lines indicated in the figures. These scans, shown next to each isophote map, quantify the radial brightness distribution across the tail  and complement the two‑dimensional morphology traced by the isophotes. The dust--loss--rate profiles associated with the two maximum particle radii are displayed in Figure~\ref{fig:dustlossrate}. These results illustrate how the nominal model parameters, together with different choices of $r_{\max}$ within the acceptable range, reproduce both the morphology of the coma, the brightness profile, and the photometric evolution of the comet in most cases. 

Before discussing the results of the model calculations, we begin by performing a series of sensitivity tests, starting from our best-fit solution for the case $r_{\max}=1$~cm. As stated above,  the number of free parameters in the model is very large, so it is not meaningful to present results obtained by varying each of them individually, especially considering that most of them are, or can be, dependent on heliocentric distance. We therefore restrict ourselves to providing a sensitivity test, examining how the isophote field change when: (a) the exponent of the $\cos(z)$ term is set to $\varepsilon$=0.5 instead of $\varepsilon$=1.0, and (b) the size--distribution function adopts a steeper power--law exponent of $\kappa=-3.8$ instead of $\kappa=-3.5$. We chose these two parameters as they are among the most influential in the space of parameters.

\begin{figure}
    \centering

    \begin{subfigure}{0.48\textwidth}
        \centering
        \includegraphics[angle=-90,trim=6cm 2cm 5cm 4cm,clip,        
        width=\linewidth]{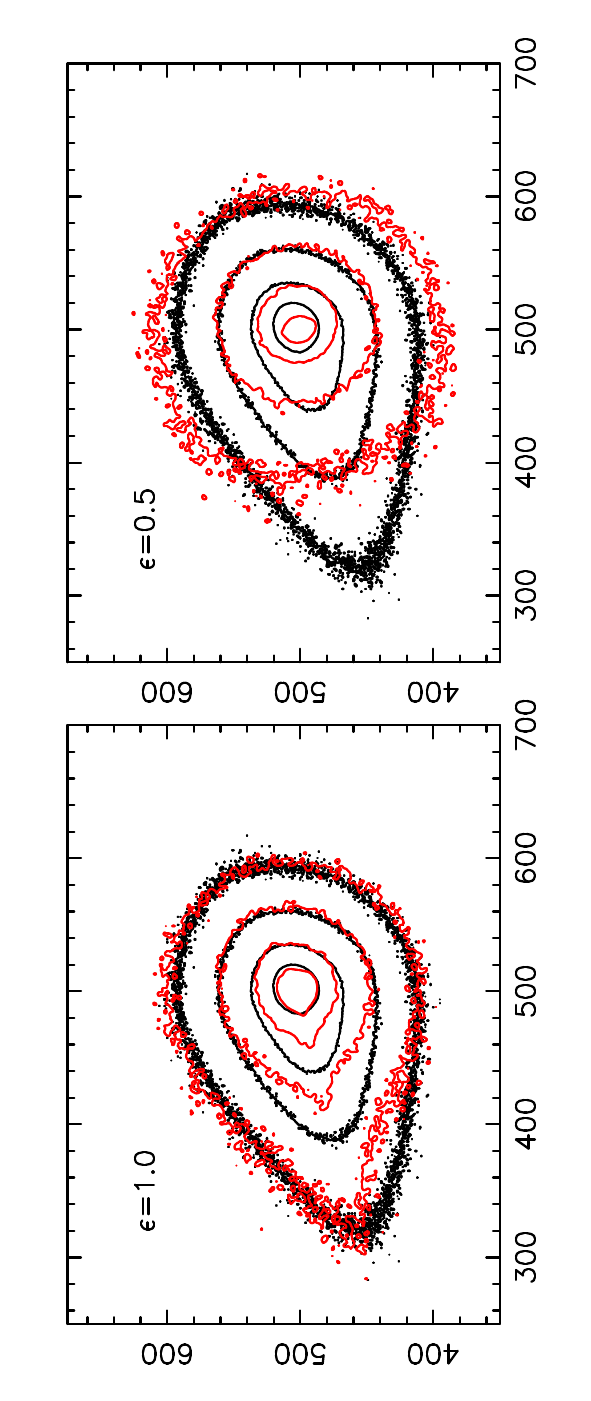}
        \caption{Left panel: Results of the model using the nominal parameters, with $r_{\max}=1$~cm and the corresponding dust--loss--rate profile of Figure~\ref{fig:dustlossrate}, applied to the image acquired on 2025-12-18 ( label \texttt{h} in Table~\ref{tab:sample_dates_post}). Right panel: Results of the model obtained by changing only the exponent of the $(\cos z)^{\varepsilon}$ term to $\varepsilon = 0.5$. Black lines show the observed isophotes, while red lines represent the model calculations. The scale is given in pixels along both the $x$- and $y$-axes.}
        \label{fig:expocos}
    \end{subfigure}

    \vspace{0.3cm}

    \begin{subfigure}{0.48\textwidth}
        \centering
        \includegraphics[angle=-90,trim=6cm 2cm 5cm 4cm,clip,          width=\linewidth]{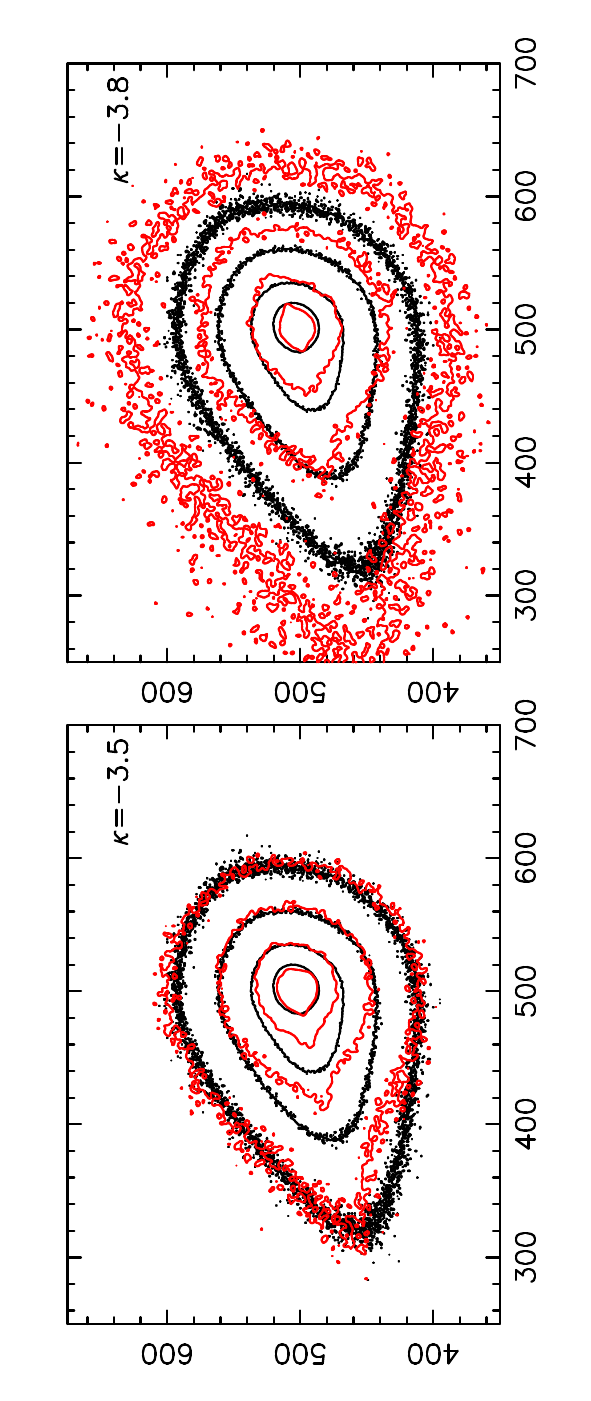}
        \caption{Left panel: As the upper left panel. Right panel: Results of the model obtained by changing the power-law size distribution index to $\kappa=-3.8$, and readjusting the dust loss rate profile to produce the best fit to the innermost isophotes.  Black lines show the observed isophotes, while red lines represent the model calculations. The scale is given in pixels along both the $x$- and $y$-axes.}
        \label{fig:kappa}
    \end{subfigure}

    \caption{Sensitivity tests to (a) the ejection velocity and (b) the power-law index of the size distribution function.}
    \label{fig:test}
\end{figure}

The results of both tests are displayed in Figure \ref{fig:test}. In the case of modifying the parameter $\varepsilon$ to a smaller value, we could not found acceptable fits to the observed isophotes, as the model gives a more symmetrical pattern around the comet photocenter, being unable to reproduce the sunward tail (see Figure \ref{fig:expocos}). When the exponent of the power-law index is set to a smaller value, e.g. $\kappa=-3.8$, we needed to readjust the dust loss rate profile by dividing it by a factor around 1.7 to fit the innermost isophotes, but then the modeled image is too bright in the outermost regions (see Figure \ref{fig:kappa}).    

As shown in Figures~\ref{fig:contours1} and \ref{fig:contours}, the fits to the observed isophotes are similar for both maximum particle sizes ($r_{\max}=1$~cm and $r_{\max}=10$~cm). However, for $r_{\max}=1$~cm the modeled isophotes display a pattern toward the solar direction near the photocenter that differs from that observed in image (l) (2026-02-04) and in subsequent images, whereas the model with $r_{\max}=10$~cm reproduces this part of the tail more accurately. In contrast, the observed light curve is better matched by the model with $r_{\max}=1$~cm, particularly along the post-perihelion branch (see Figures \ref{fig:lightcurve1} and \ref{fig:lightcurve}). The corresponding dust-loss-rate profiles differ significantly, with the $r_{\max}=10$~cm model yielding values that are larger by a factor of about 2 to 3.5 compared to those obtained for $r_{\max}=1$~cm (see Figure  \ref{fig:dustlossrate}).

Regarding particle speeds, \cite{2025MNRAS.542L.139B} estimated that millimeter-sized particles were ejected at speeds 0.01-1 m s$^{-1}$ at $r_h$=4.43 au. In our model, at that heliocentric distance, a particle of $r$=1 mm is ejected at a speed (averaged over the cosine of the zenith angle) of $\approx$ 3 m s$^{-1}$. At perihelion, it has been found an unusually low expansion velocity of coma gases, of about 370 m s$^{-1}$, which indicates that the gas flow close to the nucleus may be dominated by heavy species such as CO$_2$, or alternatively that a large fraction of the observed gas originates from the sublimation of icy grains \citep{2026arXiv260323240B}. The fastest particles ($r_{min}$=10 $\mu$m) in our model would move at mean speeds of $\approx$60 m s$^{-1}$ at perihelion. If the outgassing is dominated by CO$_2$, the drag acceleration on a dust grain of 
radius $r$ and density $\rho_{\rm d}$ follows the same free--molecular form as for 
H$_2$O, but with the larger molecular mass of CO$_2$ 
($m_{\rm g}=7.3\times 10^{-26}\,{\rm kg}$) and the corresponding CO$_2$ gas speed 
$v_{\rm g}$.  The near-surface drag acceleration is 
$a_{\rm d}\simeq (3C_{\rm D}/8)\,(\rho_{\rm g}/\rho_{\rm d}r)\,v_{\rm g}^{2}$, where $C_D$ is the drag coefficient, $\rho_g$ is the gas density, $v_g$ is the gas speed, and where   $\rho_{\rm g}(R_{\rm N}) = Q_{\rm CO_2} m_{\rm g}/(4\pi R_{\rm N}^{2} v_{\rm g})$ for a 
nucleus of radius $R_{\rm N}$.  Integrating over an acceleration length 
$L\simeq kR_{\rm N}$ yields a terminal dust speed

\begin{equation}
v_{\rm d}\sim
\sqrt{
\frac{3 k C_{\rm D}}{16\pi}\,
\frac{Q_{\rm CO_2}\,m_{\rm g}\,v_{\rm g}}{\rho_{\rm d}\,r\,R_{\rm N}}
}.
\label{eq:gaspeed}
\end{equation}

Using the appropriate parameters for 3I/ATLAS 
($r=10~\mu{\rm m}$, $\rho_{\rm d}=800~{\rm kg\,m^{-3}}$, 
$R_{\rm N}=1~{\rm km}$, $v_{\rm g}=370~{\rm m\,s^{-1}}$, 
and $Q_{\rm CO_2}=1.7\times 10^{27}~{\rm s^{-1}}$) 
\citep[this is the production rate  determined by][at $r_h=3.32$ au and therefore a lower limit compared to the perihelion rate]{2026ApJ...996L..34M}, this expression gives 
$v_{\rm d}\approx 80$--$110~{\rm m\,s^{-1}}$ for $k=3$--$5$.  
Thus a modeled dust speed of $\sim 60~{\rm m\,s^{-1}}$ for a 
$10~\mu{\rm m}$ grain remains fully compatible with CO$_2$-driven outgassing, even for the conservative lower limit on the CO$_2$ production rate.

For the other interstellar comet showing clear cometary activity, 2I/Borisov, we adopted the same size– and heliocentric–distance–dependent ejection–speed law as in Eq.~\ref{eq:speed} \citep[see][]{2020MNRAS.495.2053D}, but without the 
$\cos z$ term.  Using $\gamma=0.5$, de~Le\'on et~al.\ found a best--fit value of $v_{0}=153~\mathrm{m\,s^{-1}}$, which implies an ejection speed of $42~\mathrm{m\,s^{-1}}$ for a grain of radius $r=10~\mu\mathrm{m}$.  This value is 
smaller than, but still comparable to, the dust speed we infer for 3I/ATLAS.

Concerning the dust production rate, previous estimates have been reported by several authors, but a direct comparison with our results is difficult because most of those estimates refer to single heliocentric distances and rely on 
simplifying assumptions.  For example, \citet{2025MNRAS.542L.139B} found that at 
$r_h=4.43$~au the comet was ejecting micrometre-- to millimetre--sized particles with a mass--loss rate of $\sim$0.1--1.0~kg~s$^{-1}$.  At $r_h=3.8$~au inbound, 
\citet{2025ApJ...990L...2J} derived a dust production rate of 12--120~kg~s$^{-1}$ for particles in the 1--100~$\mu$m size range.  Observations from China’s \texttt{Tianwen--1} Mars orbiter \citep{2026arXiv260310350R} indicate particle sizes 
of a few hundred microns (from a Finson--Probstein analysis) and a dust--loss rate of $\sim$1000~kg~s$^{-1}$ at $r_h\approx1.7$~au pre--perihelion.  In addition, 
\citet{2026arXiv260301383G} estimated dust production rates from Af$\rho$ measurements, reporting an increase from $\leq 217$ to $\leq 328$~kg~s$^{-1}$ as the comet approached from $r_h=3.18$ to 2.19~au.  

All of these values are lower than our derived dust--loss rates by factors ranging from $\sim$2 up to two orders of magnitude, depending on heliocentric distance. These discrepancies arise from differences in the assumed particle--size 
distribution---with most previous studies favoring smaller grains than those retrieved in our analysis---and from the use of simplified proxies such as Af$\rho$ to infer dust production, which can underestimate the true mass--loss rate when large grains dominate the coma.

The left panel of Figure~\ref{fig:dustlossrate}  show the modeled dust loss rate of 3I/ATLAS for $r_{max}=1$ and $r_{max}=10$ cm over several hundred days around perihelion. In both cases, the activity increases gradually during the inbound leg, reaches a sharp maximum close to perihelion, and then declines rapidly afterwards. This asymmetric behavior---a slow rise followed by a faster post-perihelion decay---is typical of comets whose activity is controlled by the combined effects of volatile sublimation and thermal lag in the near-surface layers. The peak dust loss rate is confined to a relatively narrow interval around perihelion, indicating that most of the dust ejection occurs when the nucleus receives its maximum solar flux.

The right panel of Figure \ref{fig:dustlossrate} show a comparison of the modeled dust loss rate with water production rates reported by several authors, namely 
\cite{2025ApJ...991L..50X}, \cite{2026A&A...706A..43H}, 
\cite{2025RNAAS...9..242L}, \cite{2025CBET.525....1S}, \cite{2026ApJ...998L..22T}, and \cite{2026ApJ...998L..17C}. The water production increases steeply as the comet approaches perihelion
\citep[see][]{2026ApJ...998L..22T}, following a less steep post-perihelion decrease. This steep increase is not correlated to the dust production rate, that exhibits a more shallower increment. Nevertheless, we note that during the inbound leg, the gas production is dominated by CO$_2$, not H$_2$O, by a factor of 
order 5-7 \citep{2025ApJ...991L..43C, 2026ApJ..1000L..52L}, so that the total gas production rate could show a closer correlation with the dust profile, particularly for the case $r_{max}=1$ cm. As the comet approaches perihelion, the situation changes. The CO$_2$/H$_2$O ratio decreases significantly, indicating that water sublimation becomes the dominant gas source, as shown by SOHO/SWAN Lyman-$\alpha$ measurements \citep{2026ApJ...998L..22T, 2026ApJ...998L..17C}. Despite the large scatter in the measurements, the modeled dust production broadly follows the water production rate curve post-perihelion, displaying dust-to-gas ratios mostly in the range $\approx$1-4, depending on the model, the water production data set, and the specific heliocentric distance. 

Overall, the combined analysis of the modeled dust loss rate and the observed gas production rates indicates that 3I/ATLAS exhibits a strong compositional and dynamical evolution along its orbit. The transition from CO$_2$-dominated to H$_2$O-dominated activity naturally explains the changing dust-to-gas ratio and the shape of the dust production curve. The agreement between model and observations supports the interpretation that the comet's activity is controlled by the interplay between these two major volatiles, with dust release responding sensitively to the dominant gas species at each orbital phase.

\section{Conclusions}

We have carried out an extensive campaign of observations in multiple photometric bands of comet 3I/ATLAS, together with an interpretation of this data set and independent aperture photometric measurements. The analysis of the post-perihelion images with a linear--exponential model reveals a significant opposition surge of 0.1--0.4~mag, depending on the photometric band and aperture used, with a width of 1--3$^\circ$ and a linear phase coefficient in the range 0.02--0.04~mag~deg$^{-1}$. The derived phase function of the dust shows a marked contrast with previously determined phase functions for Solar System comets, suggesting that the dust particles in 3I/ATLAS may possess a higher porosity. Notwithstanding this, the \textit{apparent} colour indexes of the dust, $g-r = 0.70 \pm 0.02$ mag and $r-i = 0.22 \pm 0.05$ mag, are not significantly different from those measured in Solar System comets, and no significant color evolution is observed along the post-perihelion branch. The activity index, $n$, lies in the range 4.3--5.3, also consistent with Solar System comets.

We have performed Monte Carlo dust tail modelling of the image data set, assuming the simplest scenario of a constant power-law index $\kappa = -3.5$ and particle radii in the range $r_{\min} = 10~\mu$m to $r_{\max} = 1$--10~cm. Motivated by the sunward particle emission observed at the onset of activity, we assumed that particle speeds were proportional to the cosine of the solar zenith angle, while maintaining the customary dependence on particle size and heliocentric distance. Under these assumptions, we derived the dust-loss rates that simultaneously reproduce the photometric light curve and most of the images along the orbit. The maximum dust-loss rate occurs at perihelion and reaches $(0.5$--$1.8)\times 10^{4}$~kg~s$^{-1}$ depending on the maximum particle radius adopted in the 1 cm to 10 cm range.

Overall, the combination of a pronounced backscattering enhancement, Solar-System-like broadband colours, and dust-loss rates comparable to those of active comets near similar heliocentric distances suggests that 3I/ATLAS released dust with physical properties broadly compatible with those of Solar System comets, yet with a phase-function behaviour indicative of a higher porosity. This apparent dichotomy may reflect the predominance of highly porous aggregates in the scattering cross section, while the colour is governed by the intrinsic optical constants of the constituent materials. Future detections of interstellar comets with similarly favourable observing geometries will be essential to determine whether the scattering properties inferred for 3I/ATLAS are representative of interstellar cometary dust or instead reflect the particular evolutionary history of this object.

\section{Acknowledgements}

We are indebted to the anonymous referee for providing very useful comments and suggestions that helped us improve the paper.

We thank the amateur association \texttt{Cometas\_Obs} for providing us with aperture photometry data for 3I/ATLAS, particularly to Felipe G\'omez Pinilla. 

Based on observations made in the Transient Survey Telescope (TST) sited at the Teide Observatory of the Instituto de Astrofísica de Canarias that Light Bridges operates in the island of Tenerife, Canary Islands (Spain).
The observation time rights (DTO) used for this research on TST were consumed in the PEI "PERHELIO26". This research used storage and computing capacity in ASTRO POC's EDGE computing center at Tenerife under the form of Indefeasible Computer Rights (ICR), consumed in the PEI "PERHELIO26". The ICRs used for this research were provided by Light Bridges in cooperation with of Bechtle and LENOVO. Dr. Antonio Maudes’s insights in economics and law were instrumental in shaping the development of this work.

Based on observations collected at Centro Astronómico Hispano en Andalucía (CAHA) at Calar Alto, proposal 25B-2.2-008,  operated jointly by Junta de Andalucía and Consejo Superior de Investigaciones Científicas (IAA-CSIC).

Based on observations made with the Nordic Optical Telescope, owned in collaboration by the University of Turku and Aarhus University, and operated jointly by Aarhus University, the University of Turku and the University of Oslo, representing Denmark, Finland and Norway, the University of Iceland and Stockholm University at the Observatorio del Roque de los Muchachos, La Palma, Spain, of the Instituto de Astrofisica de Canarias. The NOT data were obtained under program ID P71-416.

FM acknowledges financial support from the grant PID2024-156684OB-I00, and from the Severo Ochoa grant CEX2021-001131-S funded by MCIN/AEI / 10.13039/501100011033.

PJG, LML and IME acknowledge support from the grant PID2021-126365NB-C21.

IME acknowledges financial support from the FPI grant PRE2022-105422 funded by MICIU/AEI/ 10.13039/501100011033 and by European Social Fund Plus (ESF+).

MRA and JL acknowledge support from the Agencia Estatal de Investigaci\'on del Ministerio de Ciencia e Innovaci\'on (AEI-MCINN) under the grants "Hydrated Minerals and Organic Compounds in Primitive Asteroids" with reference PID2020-120464GB-100 and "Hydrated Minerals and Organic Compounds in Primitive Asteroids 2" with reference PID2024-160618NB-C22. 

The Monte Carlo \texttt{FORTRAN} dust tail code \texttt{COMTAILS}, used to generate the synthetic images and the photometric data,  available at \url{https://github.com/FernandoMorenoDanvila/COMTAILS}, 
makes use of the JPL-Horizons on line ephemeris system.

\section{Data availability}

Most of the data in this paper are available from the archive images at the different observatories, and the \texttt{Cometas\_Obs} amateur association. The Monte Carlo dust tail code used to generate the synthetic images and the photometric data is available at  \url{https://github.com/FernandoMorenoDanvila/COMTAILS}. 


\bibliographystyle{mnras}
\bibliography{references} 








\bsp	
\label{lastpage}
\end{document}